\documentclass[conference]{IEEEtran}

\ifCLASSINFOpdf

\else

   \usepackage[dvips]{graphicx}

\fi

\usepackage[cmex10]{amsmath}

\usepackage{multicol}
\begin{document}
\title{Optimal Relay Power Allocation for Amplify-and-Forward Relay Networks with Non-linear Power Amplifiers}

\author{\IEEEauthorblockN{Chao Zhang, Pinyi Ren, Jingbo Peng \textdagger, Guo Wei \textdaggerdbl~ Qinghe Du and Yichen Wang}
\IEEEauthorblockA{Dept. of Information and Communication Engineering, 
Xi'an Jiaotong University, Xi'an, China\\
Email: \{chaozhang, pyren, duqinghe,wangyichen.0819\}@mail.xjtu.edu.cn\\
\textdagger\, Shanghai Research Institute, Huawei Technologies Co. LTD.\\
\textdaggerdbl\, Dept. of EEIS, University of Science and Technology of China. Hefei, China\\
Emai: jbpeng@mail.ustc.edu.cn, wei@ustc.edu.cn
}}

\maketitle
\begin{abstract}

In this paper, we propose an optimal relay power allocation of an  Amplify-and-Forward relay networks with non-linear power amplifiers. Based on Bussgang Linearization Theory, we depict the non-linear amplifying process into a linear system, which lets analyzing system performance easier. To obtain spatial diversity, we design a complete practical framework of a non-linear distortion aware receiver. Consider a total relay power constraint, we propose an optimal power allocation scheme to maximum the receiver signal-to-noise ratio. Simulation results show that proposed optimal relay power allocation indeed can improve the system capacity and resist the non-linear distortion. It is also verified that the proposed transmission scheme outperforms other transmission schemes without considering non-linear distortion.
\end{abstract}

\IEEEpeerreviewmaketitle

\section{Introduction}
Relay-assisted communication is a promising strategy that
exploits the spatial diversity available among a collection of
distributed single antenna terminals for both centralized
and decentralized wireless networks. In most relaying
networks, a two-stage relaying strategy is used \cite{IEEEhowto:laneman}\cite{Jia-Tang-relay}. In the first
stage, a source transmits and all relays listen; in the second
stage, relays cooperatively forward the source symbols to the
destination. Amplify-and-Forward (AF) and Decode-and-Forward (DF) are two most common relaying schemes \cite{IEEEhowto:laneman}. DF scheme always using cyclic redundancy check (CRC)
will cause interruptions when the relay detects errors from
the received message. If the length of source data block is significantly large, DF will incur unaffordable decoding latency \cite{IEEEhowto:latency}. AF scheme only amplify the received signal without any other processes so that little latency and complexity are introduced. On the other hand, although AF incurs noise propagation which lowers the system performance, at high power regime AF based relay network also can achieve full diversity as DF based one \cite{IEEEhowto:Jing}. Therefore, AF based relay networks recently has attracted many attentions \cite{IEEEhowto:Zhang}-\cite{IEEEhowto:shijin}. We also focus our attention on AF based relay networks. \par
On the other hand, Orthogonal Frequency Division Multiplexing (OFDM) has been absorbed into various wireless communication standards, such as Long Time Evolution (LTE), IEEE 802.16 (WiMAX) and 802.11. This is because OFDM has many benefits fit for nowadays high-speed data requirements, e.g., high spectral efficiency, resiliency to multi-path distortion and Radio-Frequency interference \cite{IEEEhowto:book}. As may be expected, OFDM technology has been applied into wireless relay networks to achieve more advantages \cite{IEEEhowto:OFDM_RElay}-\cite{IEEEhowto:rong}. Cross-Layer also has been investigated in \cite{H-Su}-\cite{X-Zhang}. 
However, it is well-known that OFDM suffers from the high Peak-to-Average Power Ratio (PAPR) problem, which always makes the power amplifier (PA) work in its non-linear region and incurs non-linear distortion for input signal \cite{IEEEhowto:book}. Consequently, PAPR also becomes a challenge to OFDM based relay networks, especially in AF based relay networks. The effect of non-linear PA for AF based relay network was investigated in \cite{IEEEhowto:BEP}. And in \cite{IEEEhowto:rec}, two receiver techniques for nonlinear
amplifier distortion compensation was proposed in an OFDM relay-assisted
cooperative communication system. So far, there are little works involved in AF based  relay networks with non-linear PA.\par
In this paper, we intend to investigate the impact of non-linear PA to the Nonlinear optimization of AF based relay networks. Particularly, the optimal relay power allocation is provided in AF based relay networks with non-linear PA. First, we introduce the Bussgang Linearization Theory which can model a non-linear process into an equivalent linear model. By this theory, we can obtain the average signal-to-noise ratio (SNR) of single source-relay-destination channel. At the destination, we also provide a practical non-linear distortion aware receiver, where Maximal-Ratio Combining (MRC) is employed to obtain spatial diversity. Besides, Optimal equivalent channel estimation through non-linear PA is also designed. Consequently, the equivalent received SNR is derived. To maximize average received SNR, we set up an optimal relay power allocation problem with total relay power constraint. After carefully analyzing, we found that if all relay power fall into a valid interval, the optimization problem is a concave function of relay power. Therefore, we finally obtain the optimal relay power allocation strategy based on Lagrange Multiplier Method. Through simulations, we can see that the proposed power allocation outperforms other power allocation strategies where non-linear distortion caused by non-linear power amplifier is not considered.
\section{System Model}
In this section, we introduce all involved models in our following analysis. First, we need identify the signal model of AF relay networks. 
\subsection{Signal Model of AF Relay Network}
\begin{figure}[t]
\centering
\includegraphics[scale=0.7]{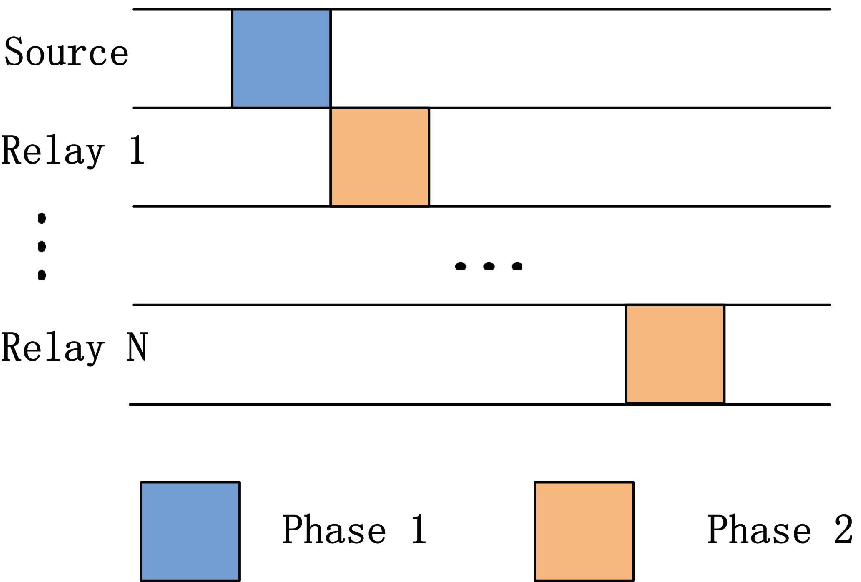}
\caption{Time slots of signal transmission}
\label{fig:timeslot}
\end{figure}
We consider a wireless network with $N$ relays, one source and one destination. Every node has a single antenna that can not transmit and receive simultaneously. Denote the channel coefficient from the source to the $i$th relay as $f_i$ and the channel coefficient from the $i$th relay to the destination as $g_i$. Assume that all $f_i$ and $g_i$ are independent complex Gaussian random variables with zero-mean and variance $\delta_{si}^{2}$ and $\delta_{id}^2$, respectively \cite{IEEEhowto:Zhang}. Note that we suppose there is no direct channel between the source node and the destination node. We further assume a block fading channel model, where channel gain stays constant during a time block and
changes from block to block. It is assumed that the instantaneous channel is unknown to the transmitting node but perfectly known at receiving node \cite{IEEEhowto:laneman}. Moreover, all relays are synchronized during relaying phase. The impact of synchronization error between relays is also beyond the scope of our discussion \cite{IEEEhowto:Jing}.\par
 The signal transmission is divided into two phases (See Fig.\ref{fig:timeslot}). During the first phase, the source broadcasts its signal $s$ to all relays, where $s$ could be a OFDM symbol \cite{IEEEhowto:BEP}. Then the $i$th relay receives
\begin{equation}
    r_i=\sqrt{P_s}f_i s+n_i
\end{equation}
where $P_s$ is the source transmit power and $n_i$ is the receiver noise which follows the complex Gaussian distribution with $n_i\sim \mathcal{CN}(0,N_0)$. Assume that there is $E\{ss^*\}$=1. After that, the relay must amplify the received signal as 
\begin{equation}\label{eq:trans_signal}
    x_i=\sqrt{\frac{P_i}{P_s\delta_{si}^2+N_0}}r_i
\end{equation}
So, there is $E\{|x_i|^2\}=P_i$. 
In previous works with a linear power amplifier, the $i$th relay would forward $x_i$ to the destination during the $i$th time-slot of the second phase (See Fig. 1). That is to say the linear amplifier seems to be transparent for the transmitted signal. However, in our consideration, $r_i$ could access the non-linear region of the front-end power amplifier with a non-vanishing probability. Model the behavior of power amplifier as a non-linear function $F_A(x)$, therefore, the signal received by the destination during the $i$th time-slot of the second phase is given by
\begin{equation}
\begin{split}
    y_i=g_iF_A(x_i)+v_i
\end{split}
\end{equation}
where $v_i$ is the Gaussian noise with $v_i\sim \mathcal{CN}(0,N_0)$. Note that we merge the non-linear effect of the source amplifier into relay amplifiers \cite{IEEEhowto:BEP}, so that the source amplifier is assumed to be linear. When the destination collects  all these $N$ forwarded signals, the transmitted signal from the source node could be detected. The receiver algorithm is discussed in section III.
\subsection{Non-linear Amplifier Model}
We assume the power amplifier (PA) of a relay is memoryless. Thus the frequency response of a PA is flat for all feed frequency. A memoryless amplifier can be determined by Amplitude to Amplitude (AM/AM) and Amplitude to Phase (AM/PM) characteristics \cite{IEEEhowto:PA_book}. Denote AM/AM transform distortion as $A_m(|x|)$ and AM/PM transform distortion as $A_p(|x|)$, then the output signal of PA is
\begin{equation} \label{eq:FA}
    F_A(x)=A_m(|x|)e^{j(arg(x)+A_p(|x|))}
\end{equation} where $\arg (x)$ means the angle of complex signal $x$.
Various amplifiers are depicted in \cite{IEEEhowto:PA_book}. 
In this paper, we choose the ideal soft-limiter amplifier (ISLA) model as an example to reveal our methodology. More practical and complicated PA models are considered in our ongoing work.
\subsubsection*{ISLA Model} 
\begin{figure}[t]
\centering
\includegraphics[scale=0.7]{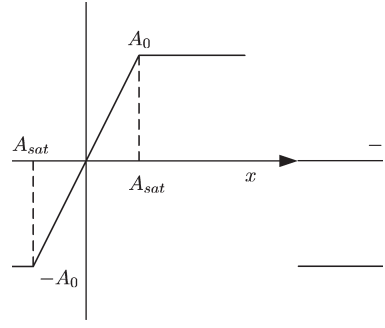}
\caption{ISLA model}
\label{fig:ISLA}
\end{figure}

\cite{IEEEhowto:PA_book}
\begin{equation}
\begin{split}
&A_m(|x|)= \left\{ \begin{array}{l}
b{|x|} + c,\;\;\;\;\;{|x|} \le {A_{sat}}\\
{A_0},\;\;\;\;\;\;\;\;\;\;\;\;{|x|} > {A_{sat}}
\end{array} \right.\\&A_p(|x|)=0
\end{split}
\end{equation} 
where $A_{sat}$ is the input saturation amplitude, $A_0$ is the output saturation amplitude and there is ${A_{sat}} = {{\left( {{A_0} - c} \right)} \mathord{\left/
 {\vphantom {{\left( {{A_0} - c} \right)} b}} \right.
 \kern-\nulldelimiterspace} b}$. Fig. \ref{fig:ISLA} shows the ISLA model. Without lose of generality, we set $c=0$, $b=1$ and $A_{sat}=A_0$. 
\subsection{Bussgang Linearization Theory}
Bussgang Linearization Theory \cite{IEEEhowto:bussgang} says that the output of a non-linear power amplifier can be expressed with a scaled version of input signal and an additive uncorrelated noise term if the input signal is a Gaussian variable:
\begin{equation}\label{eq:linear}
    F_A(x)=\alpha x + d
\end{equation} where $\alpha$ is the linear scale factor (LSF), $d$ is the non-linear distortion (NLD) and there has to be $E\{x^*d\}=0$. If the number of sub-carriers is large enough, $d$ becomes Gaussian and $d\sim \mathcal{CN}(0,\delta_d^2)$ \cite{IEEEhowto:nld_gaussian}. Thus, take into account this condition, it is easy to derive following two corollaries.
\subsubsection*{Corollary 1}
\begin{equation}
    \alpha =\frac{E\{x^*F_A(x)\}}{E\{|x|^2\}}
\end{equation}
\subsubsection*{Corollary 2}
\begin{equation}
    \delta_d^2=E\{|F_A(x)|^2\}-\alpha E\{xF_A(x)^*\}
\end{equation}
\par
\par  
Observe the power amplifier model, we just need consider AM/AM transform in following analysis. Moreover, by (\ref{eq:trans_signal}) we can see that the $x_i$ follows the Gaussian distribution so that Bussgang Linearization Theory is available in our model.\par
\subsection{Single Channel SNR Model}
Apply the linear model of (\ref{eq:linear}), then the receiver signal of the $i$th time-slot at the destination can be expressed as
\begin{equation}\label{eq:eq-signal}
    y_i= \sqrt{\frac{P_sP_i}{P_s\delta_{si}^2+N_0}}h_i s+w_i
\end{equation} where $h_i=\alpha_if_ig_i$ is the equivalent channel coefficient with $h_i\sim \mathcal{CN}(0,|\alpha_i|^2\delta_{si}^2\delta_{id}^2)$ and $w_i= \alpha_i\sqrt{\frac{P_i}{P_s\delta_{si}^2+N_0}}g_in_i+d_i+v_i$ is the equivalent receiver noise, i.e., $w_i \sim \mathcal{CN}\left(0,\frac{|\alpha_i|^2P_i\delta_{id}^2N_0}{P_s\delta_{si}^2+N_0}+\delta_{d,i}^2+N_0\right)$.
As a result, we can define the average signal-noise ration (SNR) of the $i$th received signal as 
\begin{equation}\label{eq:snr}
\Gamma_i=\frac{P_s\delta_{si}^2\delta_{id}^2}{\delta_{id}^2N_0+\rho_i(P_s\delta_{si}^2+N_0)}
\end{equation} 
where $\rho_i=\frac{\delta_{d,i}^2+N_0}{P_i|\alpha_i|^2}$. If the power amplifier is linear, i.e., $|\alpha_i|^2=1$ and $\delta_{d,i}^2=0$, then $\rho_i=N_0/P_i$ and (\ref{eq:snr}) becomes coincident with the SNR expression in \cite{IEEEhowto:Yi}. 
According to Corollary 1 and 2, there is
\begin{equation}
\rho_i=\frac{P_i(E\{|F_A(x_i)|^2\}+N_0)}{|E\{x_i^*F_A(x_i)\}|^2}-1
\end{equation}
In the Appendix, we obtain the exact expression of $\rho_i$ based on ISLA model. Therefore, we have:
\begin{equation}
\rho_i+1=\frac{\gamma_{i}\left(1-e^{-\frac{1}{\gamma_i}}\right)+\mu}{\left[\sqrt{\gamma_i}\left(1-e^{-\frac{1}{\gamma_i}}\right)+\frac{\sqrt{\pi}}{2}\mbox{Erfc}\left(\sqrt{\frac{1}{\gamma_i}}\right)\right]^2}
\end{equation} where $\mbox{Erfc}(t)=\frac{2}{\sqrt{\pi}}\int_{t}^{+\infty}e^{-{x^2}}dx$,  $\gamma_i=\frac{P_i}{A_{sat}^2}$ is the normalized relay power and $\mu=\frac{N_0}{A_{sat}^2}$ is the normalized noise power. 
\section{Optimal Relay Power Allocation}
After all $N$ time-slots of phase 2, the receiver at the destination intend to exploit $\{y_1,...,y_N\}$ to detect the signal $s$ from the source. Obviously, if we need obtain high power-efficiency, the effect of NLD should be considered in the design of receiver.
\subsection{Receiver Design}
A NLD-aware receiver was designed in \cite{IEEEhowto:rec}, where Maximal-Ratio Combining (MRC) is employed to obtain spatial diversity. However, how to compute  $\alpha_i$ and $\delta_{d,i}^2$ and how to perform MRC with received signals are not provided therein. In this paper, we also adopt the NLD-MRC idea. Different from \cite{IEEEhowto:rec}, we provide a complete procedure of NLD-MRC receiver and a practical solution to estimate the parameters of the linear model ($\alpha$ and $\delta_d^2$). Like \cite{IEEEhowto:Jing}-\cite{IEEEhowto:Yi}, we assume the statistical channel state information and power allocation strategy are known by the receiver all the time. Further more, we also assume the receiver has the knowledge of the non-linear power amplifiers equipped at relays.\par
Observe (\ref{eq:eq-signal}), to perform NLD-MRC receiving, the receiver needs know each equivalent channel coefficient $h_i$.  We intend to use the pilot symbol $s_p$ from the source node to estimate $\{h_i\}$. Denote the received pilot signal as $y_i(s_p)$ during the $i$th time-slot. Because the equivalent noise $w_i$ is Gaussian, the optimal estimation of $h_i$ is \cite{IEEEhowto:Kay}
\begin{equation}\label{eq:h}
\hat{h}_i=\left(|s_p|^2\delta_{w_i}^{-2}+\delta_{h_i}^{-2}\right)^{-1}s_p^*\delta_{w_i}^{-2}y_i(s_p)
\end{equation} where $\delta_{w_i}^2=\frac{|\alpha_i|^2P_i\delta_{id}^2N_0}{P_s\delta_{si}^2+N_0}+\delta_{d,i}^2+N_0$ and $\delta_{h_i}^2=|\alpha_i|^2\delta_{si}^2\delta_{id}^2$. Obviously, we need firstly estimate $\{\alpha_i\}$ and $\{\delta_{d,i}^2\}$ to obtain the equivalent channels $\{h_i\}$  at the NLD-MRC receiver. For this purpose, we provide a following receiving algorithm:
\rule{250pt}{1pt}\\
\begin{enumerate}
    \item \textbf{Step 1}: Use the results of Appendix to compute $E\{x_i^*F_A(x_i)\}$ and $E\{|F_A(x_i)|^2\}$.  
    \item \textbf{Step 2}: According to two corollaries in Section II, obtain the linear model parameters $\{\alpha_i\}$ and $\{\delta_{d,i}^2\}$. 
    \item \textbf{Step 3}: By (\ref{eq:h}), estimate the equivalent channel coefficients $\{\hat{h}_i\}$
    \item \textbf{Step 4}: Detect the transmitted signal with rule:  $$\arg \min_s \left\{\left|\sum_{i=1}^N\frac{y_i\hat{h}_i^*}{\sum_{i=1}^N|\hat{h}_i|^2}-s\right|^2\right\}$$. 
\end{enumerate}
\rule{250pt}{1pt}\par
In following context, we consider an ideal case where $\hat{h}_i=h_i$ to analyze the optimal system performance. In term of the NLD-MRC receiver, the average received SNR is
\[
    \Gamma_t=\sum_{i=1}^N\Gamma_i
\]
At this point, both selection combining and equal-gain combing could also be  employed herein. To achieve best performance, we only consider MRC in this paper. 
\subsection{Relay Power Allocation Problem}
Assume all power amplifiers at relays have the same model and parameters.
Observe (\ref{eq:snr}), we can see that $P_s$ and $\{P_i\}$ are the variable factors to optimize SNR. As stated in \cite{IEEEhowto:Yi}, it is difficult to optimally allocate $P_s$ and $\{P_i\}$ simultaneously. We also consider a relaxed problem to one with a fixed pre-determined $P_s$. For a relay, the actual average transmit power is $E\{|F_A(x_i)|^2\}$, which is a function of $P_i$. Assume the maximum total relay power as $P_r$, we have $\sum_{i=1}^N A_{sat}^2\gamma_i\left(1-e^{-\frac{1}{\gamma_i}}\right)\leq P_r$ by (\ref{eq:Fa}). Define the normalized maximum total relay power constraint $\gamma_r = P_r/A_{sat}^2$.  Therefore, the optimal power allocation problem is 
\begin{equation}
    \begin{split}
&\max \left\{ \Gamma_t\right\} \\
&\mbox{s.t.} \sum_{i=1}^N \gamma_i\left(1-e^{-\frac{1}{\gamma_i}}\right)\leq \gamma_r\\
&~~~~0\leq \gamma_i,~~ i=1,...,N
    \end{split}
\end{equation}
\par 
\begin{figure}[t]
\centering
\includegraphics[scale=0.55]{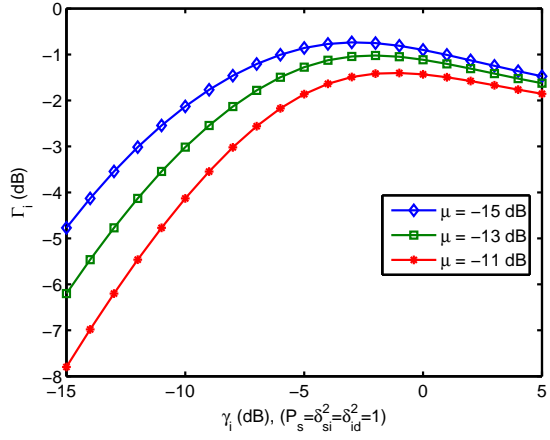}
\caption{$\Gamma_i$ versus $\gamma_i$ with all other unit parameters}
\label{fig:Gamma}
\end{figure}
\begin{figure}[t]
\centering
\includegraphics[scale=0.55]{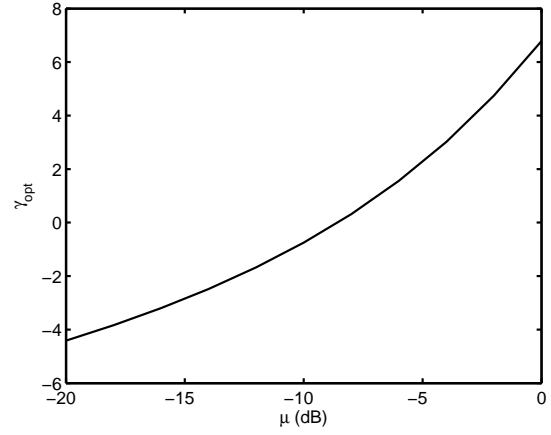}
\caption{$\gamma_{opt}$ versus $\mu$}
\label{fig:gamma_u}
\end{figure}

Observe Fig. \ref{fig:Gamma}, although $\Gamma_i$ is not a concave function of $\gamma_i$, $\Gamma_i$ could achieves the unique maximum value if $\frac{\partial \Gamma_i}{\partial \gamma_i}=0$. Then we can derive the optimal $\gamma_i$ without power constraint is the solution of
\begin{equation}\label{eq:erfc}
\frac{1}{\sqrt{\gamma_i}}=\frac{\sqrt{\pi}}{2\mu}\mbox{Erfc}\left(\frac{1}{\sqrt{\gamma_i}}\right)
\end{equation}
By (\ref{eq:erfc}), we  can also confirm that there exists an unique maximum value. We denote the solution of (\ref{eq:erfc}) as $\gamma_{opt}(\mu)$ which is independent of channel state information. Through Fig. (\ref{fig:Gamma}), we can see that if $\gamma_i$ becomes larger than $\gamma_{opt}$, $\Gamma_i$ decreases. To save power, we have no need to allocate power larger than $\gamma_{opt}$ to lower the receiver SNR.  Therefore, we just need consider $0\leq \gamma_i \leq \gamma_{opt} $ as an efficient interval \cite{IEEEhowto:jbpeng}. In addition, we plot $\gamma_{opt}$ versus $\mu$ curve in Fig. \ref{fig:gamma_u}. It shows that $\gamma_{opt}$ is a monotonic function of $\mu$. Therefore, we could set up a lookup table to show the value of $\gamma_{opt}$ for different normalized noise power.  
Furthermore, we found that $\frac{\partial^2 \Gamma_i}{\partial \gamma_i^2}\leq 0$ if $\gamma_i$ falls into the interval $0\leq \gamma_i \leq \gamma_{opt} $. 
Thus, the Hessian matrix
\begin{equation}
    \nabla^2\Gamma_t=\mbox{diag}\left\{\frac{\partial^2\Gamma_1}{\partial \gamma_1^2},\frac{\partial^2\Gamma_2}{\partial \gamma_2^2}...,\frac{\partial^2\Gamma_N}{\partial \gamma_N^2}\right\}
\end{equation}
is a negative semidefinite matrix in the efficient interval. It is obvious that $\Gamma_t$ is a concave function \cite{IEEEhowto:cov} of $(\gamma_1,\gamma_2...,\gamma_N)$ if $0\leq\gamma_i\leq \gamma_{opt}$, $i=1,...,N$. 
Therefore, we can rewrite the optimal power allocation problem as
\begin{equation} \label{eq:finalobj}
\begin{split}
&\max \left\{ \sum_{i=1}^N \Gamma_i(\gamma_i)\right\} \\
&\mbox{s.t.} \sum_{i=1}^N \gamma_i\left(1-e^{-\frac{1}{\gamma_i}}\right)\leq \gamma_r\\
&~~~~0\leq \gamma_i \leq \gamma_{opt},~~ i=1,...,N
\end{split}
\end{equation}\par
First, relax the optimization problem with only total relay power constraint.
Construct the Lagrange function 
\begin{equation}
    L=\sum_i^N \Gamma_i\left(\gamma_i\right)+\lambda \left(\sum_{i=1}^N  \gamma_i\left(1-e^{-\frac{1}{\gamma_i}}\right)-\gamma_r\right)  
\end{equation}
where $\lambda \geq 0$ is the Lagrange factor. Let $\frac{\partial L}{\partial \gamma_i}=0$, we obtain (\ref{eq:lambda}). To simplify the expression, we denote the equation of (\ref{eq:lambda}) as $\lambda=\Phi(\gamma_i,\delta_{si}^2,\delta_{id}^2)$. Hence we can express
\begin{equation}
    \gamma_i=\left(\Phi^{-1}(\lambda,\delta_{si}^2,\delta_{id}^2)\right)^{+}
\end{equation}
where $\Phi^{-1}(.)$ is the inverse function of $\Phi(.)$, $\lambda$ is a constant to meet the total relay power constraint and \[{\left( x \right)^ + } = \left\{ \begin{array}{l}
0~~~~~~~{\rm{     if  }}~~~x \le 0\\
x~~~~~~~{\rm{     else}}
\end{array} \right.\] 
Recall Fig. \ref{fig:Gamma}, $\Gamma_i$ is a monotonically increasing and concave function of $\gamma_i$ if $0\leq \gamma_i \leq \gamma_{opt}$. Therefore, the optimal solution of (\ref{eq:finalobj}) must be on the boundary \cite{IEEEhowto:Yi}\cite{IEEEhowto:cov}. As a result, the optimal relay power allocation of (\ref{eq:finalobj}) is
\begin{equation} \label{eq:solution}
    \gamma_i=\left(\Phi^{-1}(\lambda,\delta_{si}^2,\delta_{id}^2)\right)_0^{\gamma_{opt}}
\end{equation}
where 
\newcounter{mytempeqncnt}
\begin{figure*}[!t]
\normalsize
\setcounter{mytempeqncnt}{\value{equation}}
\begin{equation}\label{eq:lambda}
\lambda=\frac{P_s\delta_{si}^2\delta_{id}^2(P_s\delta_{si}^2+N_0)}{\left[\delta_{id}^2N_0+\rho_i(P_s\delta_{si}^2+N_0)\right]^2}\times \frac{e^{\frac{2}{\gamma_i}}\left(2\mu-\sqrt{\pi \gamma_i}\mbox{Erfc}\left(\frac{1}{\sqrt{\gamma_i}}\right)\right)}{2\gamma_i^2\left[-1+e^{\frac{1}{\gamma_i}}+\frac{1}{2}\sqrt{\frac{\pi}{\gamma_i}}\mbox{Erfc}\left(\frac{1}{\sqrt{\gamma_i}}\right)\right]^{3}}
\end{equation}
\hrulefill
\vspace*{4pt}
\end{figure*}
\[\left( x \right)_0^y = \left\{ \begin{array}{l}
0~~~~~~~{\rm{     if  }}~~~x \le 0\\
x~~~~~~~{\rm{     if  ~~~0}} < x \le y\\
y~~~~~~~{\rm{     if  ~~~}}x > y{\rm{ }}
\end{array} \right.\]
It is obvious that the optimal power allocation solution can be considered as an extended water-filling process \cite{IEEEhowto:Yi}, where there is a lid and bottom for each vessel. The lid is generated by $\gamma_{opt}(\mu)$. \par
The destination can estimate and collect all statistical channel state information of the relay network. Since the model of power amplifier at relays is also known by the destination, the destination can obtain the optimal relay power allocation by (\ref{eq:solution}). Through feedback link, the optimal allocation can be informed to all relays. As the optimal power allocation is based on the statistical channel state information, computing for optimal power and feedback link are not always be activated. If one relay is assigned with zero power, this relay will not forward signal during the phase 2. To enhance the spectrum efficiency, the destination could adaptively program the time-plots of phase 2 for relays with non-zero assigned power.\par

\section{Simulation Results}
\begin{figure}[t]
\centering
\includegraphics[scale=0.6]{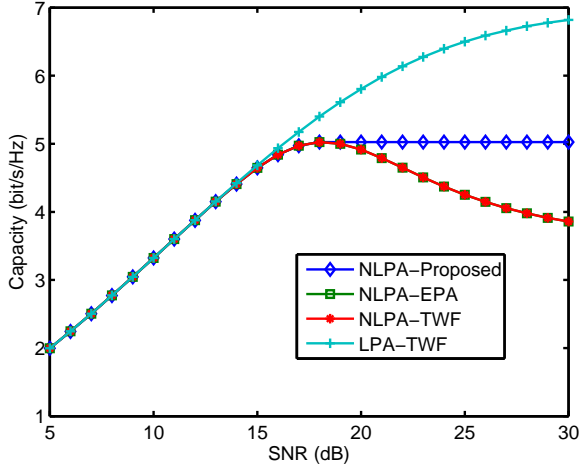}
\caption{Capacity of the symmetric relay network in four cases}
\label{fig:capacity_sym}
\end{figure}
In this section, we show simulation results to verify our analysis and to show performances of optimal power allocation. We simulate an AF network with a source, a destination and
four relay nodes ($N=4$). Without lose of generality, assign source power as $P_s=1$ and noise variance as $N_0=-15$ dB \cite{IEEEhowto:jbpeng}. Set $A_{sat}=1$, then  $\mu=-15$ dB and  Through (\ref{eq:erfc}), we have $\gamma_{opt}=-2.8507 $ dB.  Define the relay network SNR as $P_r/N_0$. Four transmission schemes with the same power constraint are compared:
\begin{figure}[t]
\centering
\includegraphics[scale=0.6]{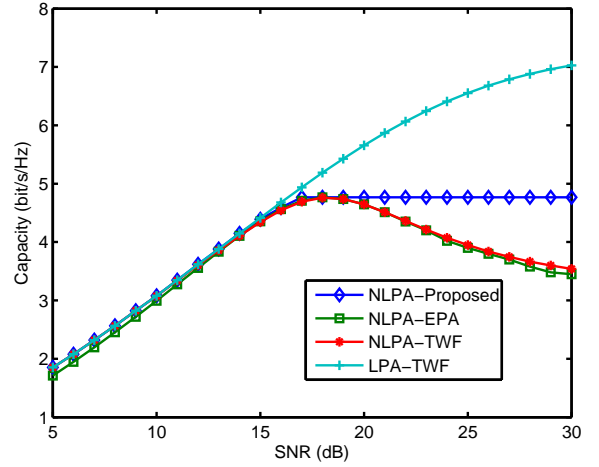}
\caption{Capacity of the non-symmetric relay network in four cases}
\label{fig:capacity_non_sym}
\end{figure}
 
\begin{itemize}
    \item With linear PA, traditional water-filling \cite{IEEEhowto:Yi} is used (LPA-TWF). 
    \item With Non-linear PA, traditional water-filling \cite{IEEEhowto:Yi} is used (NLPA-TWF). 
    \item With Non-linear PA, equal power allocation ($P_i=P_r/N$) is used (NLPA-EPA).
    \item With Non-linear PA, proposed power allocation is used (NLPA-Proposed).
\end{itemize}\par

In Fig. \ref{fig:capacity_sym}, we simulate a symmetric relay network, where each channel variance is unit, i.e., $\delta_{si}^2\sim\mathcal{CN}(0,1)$ and $\delta_{id}^2\sim\mathcal{CN}(0,1)$. Obviously, in this case,  NLPA-EPA and NLPA-TWF converge into an identical case, which is also verified by Fig. \ref{fig:capacity_sym}. As there is no NLD in LPA-TWF transmission scheme, it achieves the best performance in  all four transmission schemes. So, LPA-TWF is an ideal transmission scheme for other three schemes. To overcome NLD, a transmission scheme with non-linear PA should approach the performance of LPA-TWF as close as possible. The more close, the more ability of resisting NLD. From 5 dB to 15 dB, all transmission schemes achieve the same capacity. The reason is that the relay power is operated in the linear region of PA and the NLD derived by Bussgang theory is very slight, i.e., $\rho_i\approx N_0/P_i$. If $15$ dB $<$ SNR $< 18$dB, all other three schemes except LPA-TWF have a slower and slower capacity growing rate as SNR increases, which is because NLD starts to violate the transmitted signal from relays and the relay power gain is still larger than the harm of NLD. The higher relay transmit power is, the heavier NLD is caused. As the SNR further increases, if $SNR \geq 18$dB, the capacity of LPA-TWF still increases and NLPA-Proposed power allocation keeps a fixed capacity, but the capacities of NLPA-TWF and NLPA-EPA become to decrease. The reason is that each normalized relay power has exceeded the $\gamma_{opt}=-2.8507$ dB and relay power gain becomes smaller than the harm of NLD. Therefore, it is verified that our proposed indeed overcomes part of the NLD and outperforms other power allocation schemes without considering NLD.
\par
We also consider a non-symmetric relay network to verify our proposed power allocation in Fig. \ref{fig:capacity_non_sym}. Set each channel variance is randomly selected from the interval $(0.5,1.5)$. We average our simulation results over 1000 channel variance realizations. Compare Fig. \ref{fig:capacity_sym} and Fig. \ref{fig:capacity_non_sym}, we found that the only difference is that in both low SNR regime and high SNR regime NLPA-TWF outperforms NLPA-EPA.  In other words, the advantage of water-filling algorithm based on statistical channel state information lies in the differences between channel variances. Through entire SNR region, the proposed power allocation achieves the best transmission capacity in contrast to NLPA-TWF and NLPA-EPA and there is also a capacity gap between NLPA-proposed and NLPA-EPA in low power regime. That is to say the proposed power allocation has a better performance in non-symmetric relay networks. In summary, the proposed optimal relay power allocation can improve the system capacity and resist the NLD incurred by non-linear PA.

\section{Conclusion}
An optimal relay power allocation is proposed in this paper to maximize the receiver SNR and capacity. With the bussgang linearization theory, we convert the non-linear signal incurred by non-linear power amplifier into an equivalent linear model. A complete framework of non-linear distortion aware receiver which can perform maximal-ratio combining is proposed to achieve spatial diversity. After that, we proposed a optimal power allocation scheme with total relay power constraint. The proposed power allocation outperforms all other power allocation schemes in the presence of non-linear power amplifier and has the ability of resisting the non-linear distortion caused by non-linear power amplifier. Through simulations, we also found that proposed power allocation can achieve better performance in non-symmetric relay networks. 

\section*{Appendix}
By (\ref{eq:trans_signal}), $x_i$ is also Gaussian, i.e., $x_i\sim \mathcal{CN}(0,P_i)$. Then there is $E\{|x_i|^2\}=P_i$. Obviously, the amplitude of $x_i$ follows the Rayleigh distribution and with probability density function $$f(|x_i|)=\frac{2|x_i|}{P_i}\exp\left(-\frac{|x_i|^2}{P_i}\right)$$. 
According to (\ref{eq:FA}), there are
\begin{equation}\begin{split}
    &E\{x_i^*F_A(x_i)\}=\\
&\int_0^{+\infty} |x_i|A_m(|x_i|)e^{jA_p(|x_i|)}f(|x_i|)d|x_i|
\end{split}\end{equation} 
and 
\begin{equation}
   E\{|F_A(x_i)|^2\}=\int_0^{+\infty}A_m(|x_i|)^2f(|x_i|)d|x_i|
\end{equation}  
Consider $A_p(|x_i|)=0$, then we can deduce
\begin{equation}
    E\{x_i^*F_A(x_i)\}=E\{x_iF_A(x_i)^*\}
\end{equation}   
Apply the model of ISLA, we have
\begin{equation}
\begin{split}
&E\{x_i^*F_A(x_i)\}
=\\
&P_i\left(1-\exp\left(-\frac{A_{sat}^2}{P_i}\right)\right)+\frac{\sqrt{\pi P_i}}{2}A_{sat}\mbox{Erfc}\left(\frac{A_{sat}}{\sqrt{P_i}}\right)
\end{split}
\end{equation}
\begin{equation}\label{eq:Fa}
    E\{|F_a(x_i)|^2\}=P_i\left(1-\exp\left(-\frac{A_{sat}^2}{P_i}\right)\right)
\end{equation}

\section*{Acknowledgment}
This work was supported by New Faculty Support Program of XJTU and was
supported in part by the National Natural Science Foundation of China under
Grant No. 60832007 and the National High-Tech. Research and Development (863 Program)
Program of China under Grant No. 2009AA011801.



%

\end{document}